\documentclass[
aps,prl,
amsmath,amssymb,
superscriptaddress,
reprint]{revtex4-1}

\usepackage[T1]{fontenc}
\usepackage[utf8]{inputenc}

\usepackage{graphicx}
\usepackage[hidelinks]{hyperref}

\usepackage{color,soul}

\begin{document}

\title{Fast multifrequency measurement of nonlinear conductance}

\author{Riccardo Borgani}
\email{borgani@kth.se}
\affiliation{Nanostructure Physics, KTH Royal Institute of Technology, 10691 Stockholm, Sweden}

\author{Mojtaba Gilzad Kohan}
\affiliation{Department of Engineering Sciences and Mathematics, Luleå University of Technology, 97187 Luleå, Sweden}

\author{Alberto Vomiero}
\affiliation{Department of Engineering Sciences and Mathematics, Luleå University of Technology, 97187 Luleå, Sweden}

\author{David B. Haviland}
\affiliation{Nanostructure Physics, KTH Royal Institute of Technology, 10691 Stockholm, Sweden}

\date{17 December 2018}

\begin{abstract}
We describe a phase-coherent multifrequency lock-in measurement technique that uses the inverse Fourier transform to reconstruct the nonlinear current-voltage characteristic (IVC) of a nanoscale junction.
The method provides for a separation of the galvanic and displacement currents in the junction, and easy cancellation of the parasitic displacement current from the measurement leads.
These two features allow us to overcome traditional limitations imposed by the low conductance of the junction and high capacitance of the leads, thus providing an increase in measurement speed of several orders of magnitude.
We demonstrate the method in the context of conductive atomic force microscopy, acquiring IVCs at every pixel while scanning at standard imaging speed.
\end{abstract}

\maketitle

The sensitive measurement of small currents in nanometer-scale junctions is a central problem in modern experimental physics.
Characterization of numerous novel materials and devices, in applications ranging from topological quantum computers \cite{Lutchyn2018,Gul2018} to energy harvesting and energy conversion \cite{Si2018,Ghamgosar2018,Wedig2015,Berger2009,Liu2018,Coffey2007,Barati2017,Otnes2017}, struggles with the same basic limitations imposed by the small measurement current and the large stray capacitance of the macroscopic leads.
We describe how to circumvent these limitations using phase-coherent multifrequency lock-in measurement and inverse Fourier transform to achieve a dramatic improvement in the speed of measurement, or alternatively, in the signal-to-noise ratio at the same measurement speed.
In addition, our frequency-domain approach allows for active cancellation of parasitic current due to the lead capacitance and it provides for unambiguous separation of the galvanic and displacement currents flowing in the nanoscale junction.

One area where this improvement is particularly useful is scanning probe microscopy (SPM), where a measurement of the nonlinear current-voltage characteristic (IVC) is desired at each tip location.
In scanning tunneling microscopy (STM) the IVC allows for mapping energy dependence of the local density of electronic states \cite{Galvis2018}.
In conducting atomic force microscopy (AFM) it can be used to map energy-conversion efficiency in photoactive nanocomposite materials \cite{Mikulik2017}.
Due to the aforementioned limitations, present-day SPM has two basic modes of operation.
In imaging mode the current is measured while scanning the surface with a constant tip-sample voltage, quickly generating an image but with only limited information.
Multiple scans at different bias are required to get the full IVC, greatly increasing measurement time and introducing problems due to instrument drift and tip wear.
In spectroscopic mode the IVC is recorded at each tip position, but the voltage must be swept slowly so as to minimize displacement current in the parallel capacitance of the measurement leads.
This large background current puts a limit on the achievable gain and sensitivity of current measurement, and the slow sweep greatly limits the speed of the scan, or equivalently spatial resolution in a given measurement time.
We demonstrate how to bridge the gap between these two modes of SPM, achieving complete electrical characterization at scanning speeds characteristic to imaging mode.

Pioneering work in this general direction used high-speed acquisition to capture data while scanning at imaging speeds \cite{Somnath2018}.
The large data sets \cite{Belianinov2015} (several GBytes/scan) were subsequently analyzed with advanced filtering methods based on statistical inference, separately reconstructing galvanic current and capacitance at each image pixel.
However, this `big data' approach is computationally expensive, requiring several hours of analysis on the fastest computers.

In contrast, we take a deterministic (physical) approach which exploits our knowledge of the periodicity and phase of the applied bias.
Because we know the period of the drive waveform, frequency-domain representation of the nonlinear response efficiently rejects noise and compresses data to a manageable size for storage.
The inverse Fast Fourier Transform (FFT) algorithm makes the analysis computationally efficient and easily performed in real-time on a notebook computer.
Apart from the physical model of the measurement, our reconstruction does not require assumptions about the functional form of the nonlinear galvanic current or the bias dependence of the junction capacitance.

Below we describe our method using conductive AFM as an example, but the general idea has broad applicability in many areas of experimental physics where, with the recent development of multifrequency digital lock-in amplifiers, its implementation has been greatly simplified.

\textit{Theory} --
\begin{figure}
    \centering
    \includegraphics[width=\columnwidth]{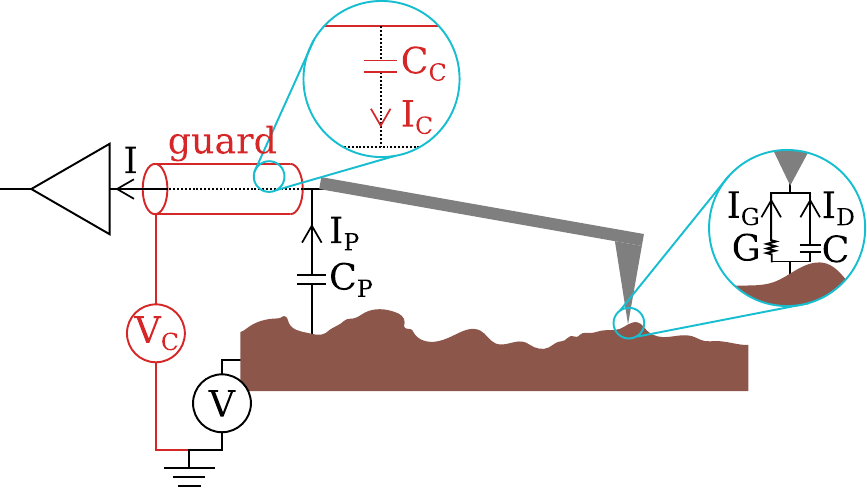}
    \caption{\textbf{Experimental setup and electrical connections.}
    The conductance $G$ and the capacitance $C$ model the tip-sample junction, $C_\mathrm{P}$ the parasitic capacitance, $C_\mathrm{C}$ the capacitance of the coaxial cable.
    $V$ is the tip-sample bias voltage, and $V_\mathrm{C}$ is the compensation drive to the guard of the coaxial cable.
    $I$ is the total current at the input of the transimpedance amplifier.
    The compensation circuit is shown in red.
    }
    \label{fig:scheme}
\end{figure}
Figure~\ref{fig:scheme} shows a schematic representation of an SPM measurement and its equivalent circuit.
When a time-dependent bias voltage $V(t)$ is applied between the SPM tip and a conductive sample, the total current measured at the amplifier is
\begin{equation}
\begin{split}
    I &= G(V)V + C(V)\dot{V} + C_\mathrm{P}\dot{V} + C_\mathrm{C}\dot{V}_\mathrm{C}\\
    &\equiv I_\mathrm{G} + I_\mathrm{D} + I_\mathrm{P} + I_\mathrm{C}
    .\label{eq:tot_current}
\end{split}
\end{equation}
We call the terms $I_\mathrm{G}$, $I_\mathrm{D}$, $I_\mathrm{P}$ and $I_\mathrm{C}$ the galvanic, displacement, parasitic and compensation currents, respectively.
The bias and its time derivative $\dot{V}$ are known signals.
The conductance $G$ and capacitance $C$ between the tip and the sample are the desired quantities in our physical model.
We emphasize that both the conductance $G(V)$ and the capacitance $C(V)$ can be nonlinear functions of $V$.
$C_\mathrm{P}$ is a distributed parasitic capacitance in parallel with the junction, modeling the capacitive contributions of the probe body, cables, and measurement electronics.
Good electrical design with proper guarding can minimize $C_\mathrm{P}$.
However, when measuring current in the subpicoampere range, the residual $I_\mathrm{P}$ due to the unguarded $C_\mathrm{P}$ can easily saturate the transimpedance amplifier, limiting its gain and thus sensitivity.
We compensate for this parasitic current as follows.

With the tip lifted away from the sample surface (tens to hundreds of micrometers), the first two terms in Eq.~(\ref{eq:tot_current}) are both zero.
The measured current is then $I_\mathrm{lift} = C_\mathrm{P}\dot{V} + C_\mathrm{C}\dot{V}_\mathrm{C}$,
where $V_\mathrm{C}$ is applied to the guard.
We obtain the value for $C_\mathrm{P}$ by measuring $I_\mathrm{lift}$ when applying a known $\dot{V}$ and keeping $V_\mathrm{C}$ at zero, and similarly for $C_\mathrm{C}$:
\begin{equation}
   C_\mathrm{P} = \left.\frac{I_\mathrm{lift}}{\dot{V}}\right|_{V_\mathrm{C}=0},
   \quad\label{eq:C_P}
    C_\mathrm{C} = \left.\frac{I_\mathrm{lift}}{\dot{V}_\mathrm{C}}\right|_{V=0}.
\end{equation}
Once $C_\mathrm{P}$ is obtained from Eq.~(\ref{eq:C_P}), the parasitic displacement current $I_\mathrm{P}$ can be \emph{passively} compensated for by subtracting it from the measured current.
In order to not limit the amplifier gain, however, it is far better to \emph{actively} compensate for $I_\mathrm{P}$ by nulling it before it reaches the amplifier:
with $C_\mathrm{C}$ from Eq.~(\ref{eq:C_P}), we apply a $V_\mathrm{C}$ that exactly cancels the contribution from the parasitic capacitance,
$V_\mathrm{C}(t) = -(C_\mathrm{C}/C_\mathrm{P})V(t)$.
Because $C_\mathrm{P}$ is at the millimeter to meter scale (probe holder, cables, and measurement electronics), its value is not changing significantly between the lifted and scanning positions of the SPM tip.
The compensation voltage $V_\mathrm{C}$ is therefore constant while scanning.

To measure the nonlinear galvanic and displacement currents, we apply a time-dependent sample bias of the form $V(t)=V_\mathrm{AC} \cos(\omega_1 t)$.
Assuming the junction conductance and capacitance are analytic functions of voltage, they both share the same time periodicity as $V$ and can therefore be written as a Fourier series: $G[V(t)]=\sum_{m} g_{m} \cos(m \omega_1 t)$ and $C[V(t)]=\sum_{m} c_{m} \cos(m \omega_1 t)$.
The two components of the current are:
\begin{eqnarray}
    I_\mathrm{G}(t) &=& G(t) V_\mathrm{AC}\cos(\omega_1 t)
    =\sum_{k=0}^{+\infty} I_{\mathrm{G}k} \cos(k \omega_1 t),
     \label{eq:Ig}\\
    I_\mathrm{D}(t) &=& -C(t)\omega_1V_\mathrm{AC}\sin(\omega_1 t) = \sum_{k=0}^{+\infty} I_{\mathrm{D}k} \sin(k \omega_1 t),
     \label{eq:Ic}
\end{eqnarray}
where the $\{I_{\mathrm{G}k}\}$ and $\{I_{\mathrm{D}k}\}$ are real constants.
Equations~(\ref{eq:Ig}) and~(\ref{eq:Ic}) show that the two current contributions are easily distinguishable: the galvanic current is in phase with $V$, and the displacement current is in phase with $\dot{V}$.
We can therefore obtain the galvanic and displacement currents from the real and imaginary part, respectively, of the Fourier transform of the measured compensated current:
\begin{equation}
    \hat{I}_\mathrm{G}(\omega)=\Re\left[\hat{I}(\omega)\right],
    \quad
    \hat{I}_\mathrm{D}(\omega)=\mathrm{i}\Im\left[\hat{I}(\omega)\right].
    \label{eq:Igdw}
\end{equation}

\textit{Experimental results} --
\begin{figure*}
    \centering
    \includegraphics[width=\textwidth]{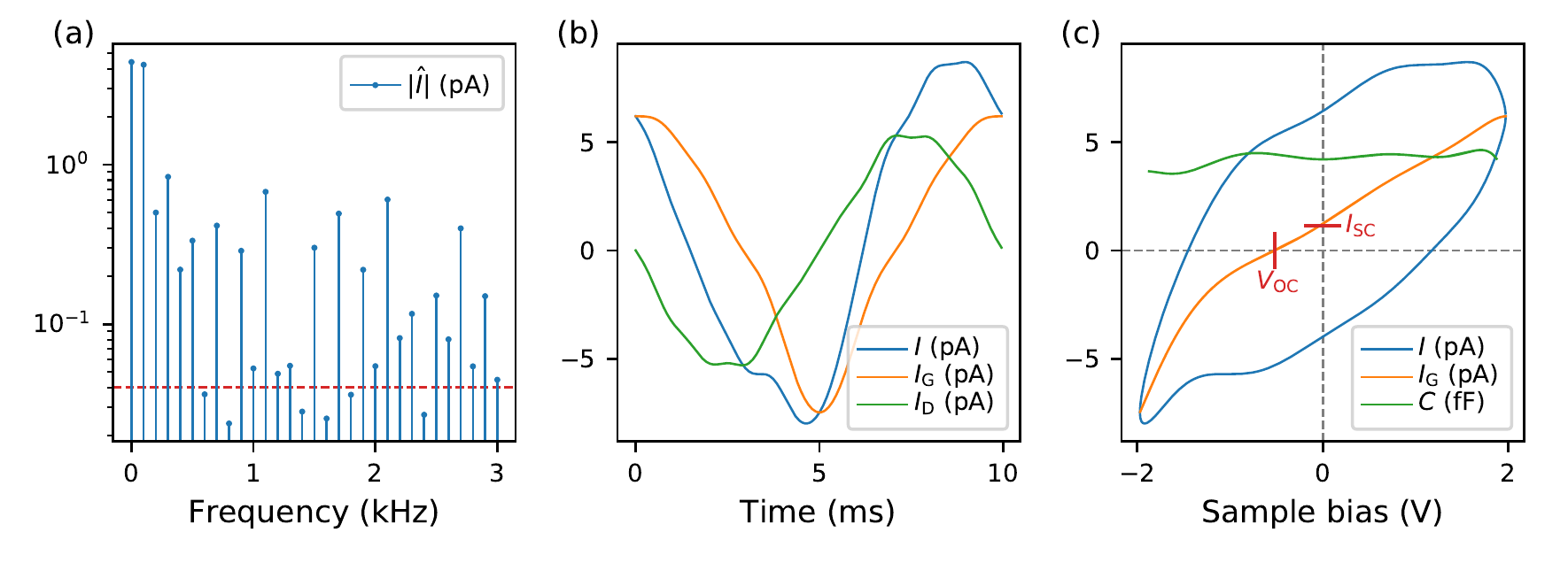}
    \caption{\textbf{Current-voltage characteristics on a photo-active polymer blend.}
    Measurement under white-light illumination.
    (a), amplitude of the measured current at 31 frequencies, phase is also measured but not shown.
    The red dashed line is the calibrated noise level.
    (b), total, galvanic and displacement currents as a function of time obtained from the current spectrum by inverse Fourier transform of Eq.s~(\ref{eq:Igdw}).
    (c), reconstructed currents and junction capacitance vs. voltage.
    The loop in the total current $I(V)$ is due to the junction capacitance.
    The galvanic current $I_G(V)$ does not show such a loop.
    The junction capacitance $C$ is nearly constant, as expected.
    }
    \label{fig:IV}
\end{figure*}
\begin{figure*}
    \centering
    \includegraphics[width=\textwidth]{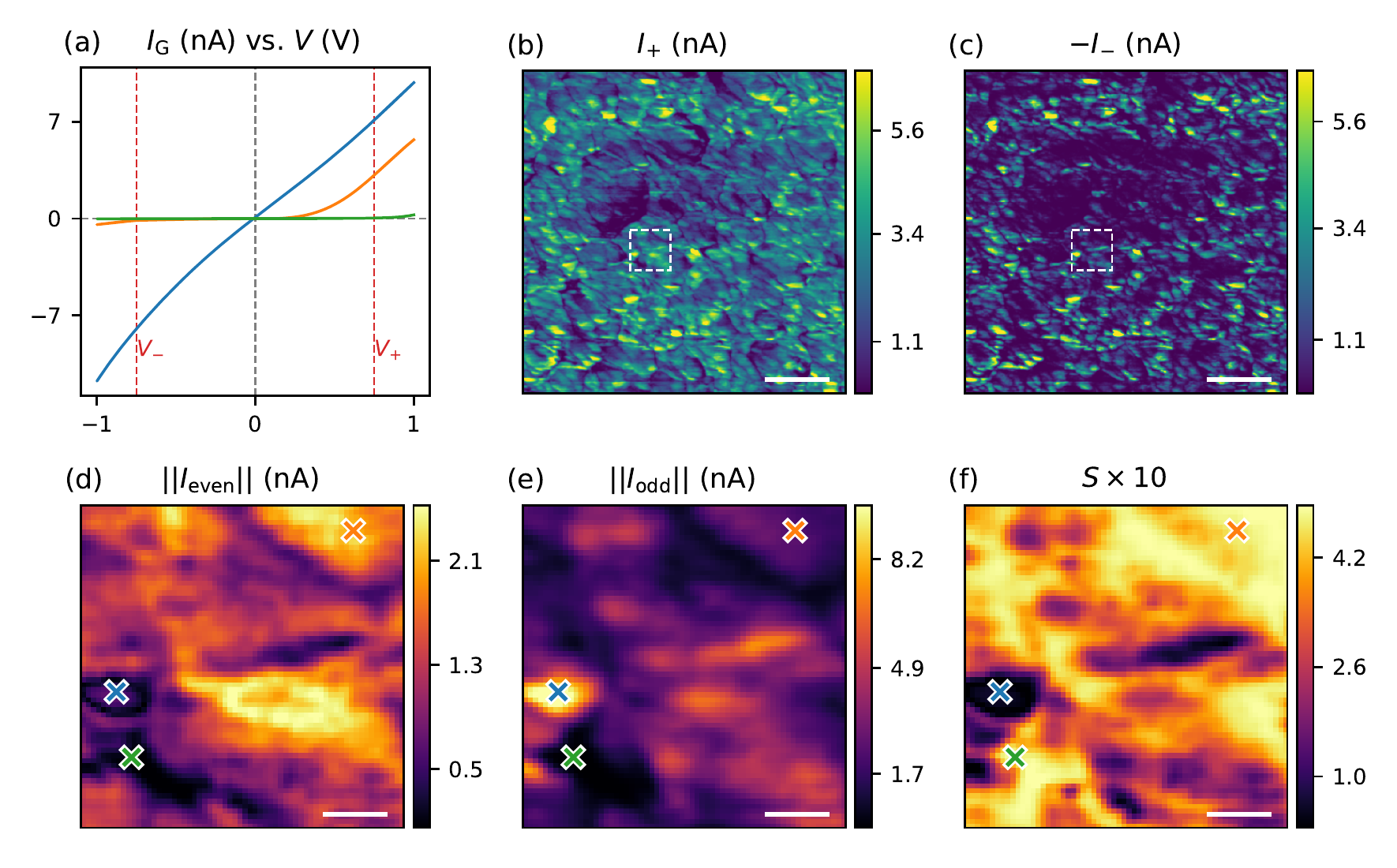}
    \caption{\textbf{Measurements on a thin-film all-oxide p/n junction.}
    (a), representative IVCs on different locations of the sample.
    Maps of: (b), current $I_\mathrm{+}$ at forward bias $V_\mathrm{+} = 0.75~\mathrm{V}$; (c), current $I_\mathrm{-}$ at reverse bias $V_\mathrm{-} = -0.75~\mathrm{V}$; (d), even-symmetry current $||I_\mathrm{even}||$; (e), odd-symmetry current $||I_\mathrm{odd}||$; (f), symmetry parameter $S = ||I_\mathrm{even}|| / (||I_\mathrm{even}|| + ||I_\mathrm{odd}||)$.
    The maps in (d), (e) and (f) are on a zoomed area marked by the white dashed boxes in (b) and (c).
    The white scale bars are $200~\mathrm{nm}$ in (b-c), and $25~\mathrm{nm}$ in (d-f).
    The crosses in (d-f) mark the positions were the IVCs of corresponding colors are acquired.
    }
    \label{fig:combi_luleo}
\end{figure*}
Figure~\ref{fig:IV} demonstrates the technique with conducting AFM, measuring the IVC between a Pt coated AFM tip and an organic solar cell consisting of a TQ1:T10 polymer blend on a PEDOT:PSS/PEI transparent electrode.
The Fourier components of the current $\hat{I}(\omega)=\hat{I}_\mathrm{G} + \mathrm{i}\hat{I}_\mathrm{D}$ are measured simultaneously at 31 harmonics of $\omega_1$ [Fig.~\ref{fig:IV}(a)] with a multifrequency lock-in amplifier \cite{IMP_AB}.
Using the inverse Fourier transform we obtain the total current $I$ (blue), galvanic current $I_\mathrm{G}$ (orange) and displacement $I_\mathrm{D}$ (green), as shown in Fig.~\ref{fig:IV}(b).

Finally in Fig.~\ref{fig:IV}(c) we plot $I_\mathrm{G}(t)$ versus the applied $V(t)$ to obtain the IVC, free of any capacitive contribution (orange curve).
The organic photovoltaic material in Fig.~\ref{fig:IV} was measured under white light illumination, where the galvanic IVC shows typical photodiode response, with an offset current at zero applied voltage (short-circuit current $I_\mathrm{SC}$) and a finite voltage required to obtain zero current (open-circuit voltage $V_\mathrm{OC}$).
For comparison, we also plot the IVC that one would obtain from the total current (blue curve), \textit{i.e.} without separation of the galvanic and displacement contributions.
The junction capacitance produces a big hysteresis loop which completely masks these important features of the galvanic IVC.
In standard conducting AFM, this hysteresis loop would be even larger as the parasitic capacitance $C_\mathrm{P}$, eliminated by active cancellation, is orders of magnitude larger than the junction capacitance $C$.

The ability to separate the junction's displacement current from its galvanic current allows us to plot the junction capacitance $C(V)= I_\mathrm{D}(t) / \dot{V}(t)$ versus $V(t)$, shown by the green curve in Fig.~\ref{fig:IV}(c).
Note that this capacitance is ideally that of the junction itself as the large parallel capacitance is removed when compensating for $I_\mathrm{P}$.
However, some residual parallel capacitance local to the tip may be included, depending the exact geometry of the tip and surface when nulling $I_\mathrm{P}$.
The ability to measure the voltage dependence of the tip-sample capacitance in parallel with the galvanic IVC is an exciting feature of this technique, which may be useful to investigate the screening length in two-dimensional electron gases \cite{Giannazzo2009} or quantum capacitance in low-dimensional devices \cite{Ilani2006}.

Acquiring the full IVC at every pixel of an AFM scan allows for the rapid mapping of interesting electrical properties such as $I_\mathrm{SC}$ and $V_\mathrm{OC}$.
Due to the computational efficiency of the reconstruction, quantities such as the maximum-power point or fill-factor can be evaluated and displayed in real time during the scan (see maps in Supplemental material) and these quantities can be correlated with the topography to investigate structure-property relationships.
As a demonstration of the resolution enabled by this measurement method, we analyze another sample consisting of a copper oxide $\mathrm{CuO_2}$ / zinc oxide $\mathrm{ZnO}$ thin-film p/n junction deposited on a fluorine-doped tin oxide (FTO) conducting glass.
Such composite oxide materials find application in sensing \cite{Kwon2018}, energy conversion \cite{Concina2017,Ruhle2012} and lighting \cite{Kim2017}, and their nanoscale characterization is needed for understanding their functionality.

Figure~\ref{fig:combi_luleo}(a) shows that the junction of the two semiconductors presents a typical diode-like IVC, with current flowing only for positive applied bias (orange curve).
However, some areas of the sample show a symmetric IVC (blue curve) indicating a damage in the $\mathrm{CuO_2}$ or a resistive phase of the composite material.
Other areas show zero current in both bias directions (green curve), indicating much smaller area of contact between the tip and the sample, or perhaps an insulating phase of material.
To analyze the spatial distribution of these three classes of IVC, we plot the current $I_\mathrm{+}$ at fixed forward bias $V_\mathrm{+} = 0.75~\mathrm{V}$ and the current $I_\mathrm{-}$ at fixed reversed bias $V_\mathrm{-} = -0.75~\mathrm{V}$.
We note that the maps $I_\mathrm{\pm}$ can be obtained with standard conductive AFM, by performing multiple scans with different applied bias $V_\mathrm{\pm}$.
Here, however, current maps for any bias $-V_\mathrm{AC}<V<V_\mathrm{AC}$ can be calculated from the measured data, as the whole IVC is acquired in a single scan.

In the forward bias image [Fig.~\ref{fig:combi_luleo}(b)] both diode-like and resistor-like areas have high current and appear as bright spots, while areas with no current are dark.
In the reverse bias image [Fig.~\ref{fig:combi_luleo}(c)], the resistor-like areas stand out as bright spots while the diode-like and no-current areas remain dark.
To more precisely distinguish the diode-like areas, we focus on a small region of these images (white dashed lines) and we analyze the symmetry properties of the acquired IVCs: resistor-like curves have a clear odd symmetry around zero [Fig.~\ref{fig:combi_luleo}(e)], and regions with diode-like curves have a stronger even-symmetry component [Fig.~\ref{fig:combi_luleo}(d)].
The two images are combined in Fig.~\ref{fig:combi_luleo}(f) using the symmetry parameter~$S$, which is zero for a purely odd curve and one for a purely even curve (for more details see Supplemental material).

The fine detail in these images is possible due the very high density of measured IVCs, like the ones of Fig.~\ref{fig:combi_luleo}(a) which were acquired at $1000~\mathrm{pixels/sec}$, yielding a trace-retrace image with 512x512 resolution (524,288 IVCs, file size 170 MB) in less than 9 minutes.
With conductive AFM in spectroscopic mode, one would typically sweep the bias in $1-10$ seconds, giving a total measurement time of $6-60$ days for the same spatial resolution.
The analysis of the full image requires about 4 seconds on a laptop computer, and it is therefore easily performed in real-time, \textit{i.e.} while the AFM is scanning.

The pixel rate is set by the measurement bandwidth $\Delta \omega$ which has a maximum value of $\Delta \omega=\omega_1$.
The choice of $\omega_1$ in turn determines how many harmonics can be resolved within the bandwidth of the transimpedance amplifier, and thereby the sharpness of features in the reconstructed IVC. 
The measurement bandwidth also sets the signal-to-noise ratio (noise $\propto \sqrt{\Delta \omega}$).
Figure~\ref{fig:IV}(a) shows the noise level $0.04~\mathrm{pA}$ ($4~\mathrm{fA}/\sqrt{\mathrm{Hz}})$, which allowed for $n=10$ harmonics within the amplifier bandwidth for $\omega_1/2 \pi = 100~\mathrm{Hz}$.
In Fig.~\ref{fig:combi_luleo} the noise level is $1.3~\mathrm{pA}$ ($40~\mathrm{fA}/\sqrt{\mathrm{Hz}})$, with $\omega_1/2 \pi = 1~\mathrm{kHz}$ and $n=50$.
When $n$ harmonics are measured in a given time (inverse measurement bandwidth) a factor $n$ improvement of measurement speed is achieved at the same signal-to-noise ratio, in comparison to traditional time-domain methods.

An important consideration with the method described here is that $\omega_1$, its harmonics, and the sampling frequency used by the digital multifrequency lock-in amplifier must both be integer multiples of the measurement bandwidth $\Delta \omega$.
This `tuning' eliminates Fourier leakage in the harmonic spectrum, ensuring that the amplitude and phase of all the harmonics are measured coherently.
The nonlinear information of the IVC is then coded in the harmonics and revealed by simple inverse Fourier transform.

\textit{Conclusions} --
We described and demonstrated a new measurement paradigm for capturing nonlinear current-voltage characteristics from weak and noisy signals.
Using one stable reference oscillation for phase-sensitive detection of many harmonics, we achieved frequency-domain multiplexing of the information contained in the nonlinear IVC.
Another important advantage of our frequency-domain approach is that we can easily extract the separate contributions of the galvanic and displacement currents in the measured total current.
The frequency-domain data is an optimally compressed representation of the nonlinear response and it is computationally efficient to reconstruct the IVC using the inverse FFT algorithm.
The frequency domain approach also provides for simple cancellation of the large parasitic current due to the stray capacitance of the leads, allowing for larger gain without saturation of the current amplifier.
Together these advantages allow for greatly enhanced measurement speed, compact data storage, and real-time feedback while measuring.

We used scanning probe microscopy to demonstrate the power of the method, reconstructing the full IVC at every pixel of a conducting AFM image, without compromising the scanning speed.
With the IVC at every pixel we constructed \textit{a posteriori} images that highlight the interesting figures of merit for electric transport.
With standard conductive AFM the set of bias voltages is decided \textit{a priori}, before performing multiple scans.
During the long measurement, instrument drift and tip wear inhibit reliable correlation between tip position and measured electrical properties.
The multifrequency approach provides a general solution to a very general and common problem in nanotechnology and its easy implementation has recently been made possible with the advent of tuned multifrequency digital lock-in amplifiers.

\begin{acknowledgments}
The authors acknowledge O. Ingenäs and Y. Xia for providing the organic solar cell sample.
DBH and RB acknowledge financial support from the Swedish Research Council (VR), and the Knut and Alice Wallenberg Foundation.
AV and MGK acknowledge financial support from the Kempe Foundation and the Knut and Alice Wallenberg Foundation.
\end{acknowledgments}

%

\end{document}